\documentclass[a4paper]{PoS}

\usepackage{textcomp}
\usepackage{amsfonts} % AMS
\usepackage{amssymb} % AMS
\usepackage{amsmath} % AMS
\usepackage{slashed}
\usepackage{graphicx}
\usepackage{pifont}
\usepackage{multirow,lscape}
\usepackage{array}
\usepackage{longtable}
\usepackage[FIGBOTCAP,TABBOTCAP,tight]{subfigure}
\usepackage{verbatim}
\usepackage{bm}
\usepackage{comment}
\usepackage{color}
\usepackage{url} %%% URL
\usepackage{caption}

\graphicspath{{./fig/}}

\renewcommand{\Re}{\mathrm{Re}}
\renewcommand{\Im}{\mathrm{Im}}

\providecommand{\MeV}{\,\mathrm{MeV}}

\newcolumntype{C}[1]{>{\centering\arraybackslash}p{#1}}

\newcommand{\BK}{\hat{B}_{K}}

\newcommand{\eps}{\varepsilon}

\newcommand{\epsK}{\varepsilon_{K}}

\title{Status report on $\varepsilon_K$ with lattice QCD inputs}

\ShortTitle{$\varepsilon_K$ in lattice QCD}

\author{Jon A.~Bailey, \speaker{Weonjong Lee},
  Jaehoon Leem, and Sungwoo Park \\
  Lattice Gauge Theory Research Center, CTP, and FPRD, \\
  Department of Physics and Astronomy, \\
  Seoul National University,
  Seoul 08826, South Korea\\
  E-mail: \email{wlee@snu.ac.kr} }

\author{Yong-Chull Jang \\
  Los Alamos National Laboratory, \\
  Theoretical Division T-2, MS B283, \\
  Los Alamos, New Mexico 87545, USA \\
  E-mail: \email{integration.field@gmail.com} }

\author{SWME Collaboration}

\abstract{ We report the current status of $\epsK$, the indirect CP
  violation parameter in the neutral kaon system, evaluated using the
  lattice QCD inputs.
  We use lattice QCD to fix $\BK$, $\xi_0$, $\xi_2$,
  $|V_{us}|$, $m_c(m_c)$, and $|V_{cb}|$.
  Since Lattice 2015, FLAG updated $\BK$, exclusive $V_{cb}$ has been
  updated with new lattice data in the $\bar{B}\to D\ell\nu$ decay channel, and
  RBC-UKQCD has updated $\xi_0$ and $\xi_2$.
  Our preliminary results show that the standard model evaluation
  of $\epsK$ with exclusive $|V_{cb}|$ (lattice QCD inputs) has
  $3.2\sigma$ tension with the experimental value, while that of
  $\epsK$ with inclusive $|V_{cb}|$ (heavy quark expansion) shows no
  tension.
 }

\FullConference{34th annual International Symposium on Lattice Field Theory\\
		24-30 July 2016\\
		University of Southampton, UK}

\begin{document}

\section{Introduction}

This paper is a follow-up and update of our previous paper \cite{
  Bailey:2015tba, Bailey:2015frw}.
In the standard model, the indirect CP violation parameter of the 
neutral kaon system $\epsK$ is
\begin{align}
  \label{eq:epsK_def}
  \epsK
  & \equiv \frac{\mathcal{A}(K_L \to \pi\pi(I=0))}
              {\mathcal{A}(K_S \to \pi\pi(I=0))}
  \nonumber \\
  & =  e^{i\theta} \sqrt{2}\sin{\theta}
  \Big( C_{\eps} \hat{B}_{K} X_\text{SD}
  + \frac{ \xi_{0} }{ \sqrt{2} } + \xi_\text{LD} \Big) 
   + \mathcal{O}(\omega\eps^\prime)
   + \mathcal{O}(\xi_0 \Gamma_2/\Gamma_1) \,, 
\end{align}
where $C_{\eps}$ is a well-known coupling, and $X_\text{SD}$
is the short distance contribution from the box diagrams.
Master formulas for $C_{\eps}$, $X_\text{SD}$, $\xi_0$, and
$\xi_\text{LD}$ are given in Ref.~\cite{Bailey:2015tba}.

Since Lattice 2015, there have been major updates of lattice QCD
inputs such as $V_{cb}$, $\BK$, $\xi_0$, and $\xi_2$.
Hence, it is time to update the current status of $\epsK$.

\section{Input parameter $|V_{cb}|$}
%
%
% Table of input parameters
%
\begin{table}[ht]
  \begin{minipage}[b]{0.44\linewidth}
  \centering
  \renewcommand{\arraystretch}{1.25}
  \resizebox{1.0\textwidth}{!}{
    \begin{tabular}{lll}
      \hline\hline
      Decay mode & $|V_{cb}|$ & Ref. 
      \\ \hline
      $\bar{B}\to D^*\ell\bar{\nu}$ & $39.04(49)(53)(19)$ & 
      \cite{Bailey2014:PhysRevD.89.114504}
      \\ \hline
      $\bar{B}\to D\ell\bar{\nu}$ & $40.7(10)(2)$ & \cite{DeTar:2015orc}
      \\ \hline
      ex-combined & $39.62(60)$ & this paper
      \\ \hline
      $\bar{B}\to X_c\ell\bar{\nu}$ & 42.00(64) & \cite{Gambino:2016jkc}
      \\ \hline\hline
      Decay mode & $|V_{ub}|$ & Ref. 
      \\ \hline
      $\bar{B}\to \pi\ell\bar{\nu}$ & $3.70(14)$ & 
      \cite{ Lattice:2015tia, Flynn:2015mha}        
      \\ \hline
      $\bar{B}\to X_u\ell\bar{\nu}$ & $4.45(16)(22)$ & \cite{Amhis:2014hma}
      \\ \hline\hline
      Decay mode & $|V_{ub}/V_{cb}|$ & Ref. 
      \\ \hline
      $\Lambda_b\to \Lambda_c\ell\bar{\nu}$ & $0.083(4)(4)$ & 
      \cite{ Detmold:2015aaa}        
      \\ \hline\hline
    \end{tabular} 
  } %%% \resizebox
  \caption{Results of $|V_{cb}|$ and $|V_{ub}|$.}
  \label{tab:Vcb-Vub}
  \end{minipage}
  \hfill
  \begin{minipage}[b]{0.56\linewidth}
    \centering
    \includegraphics[width=\textwidth]{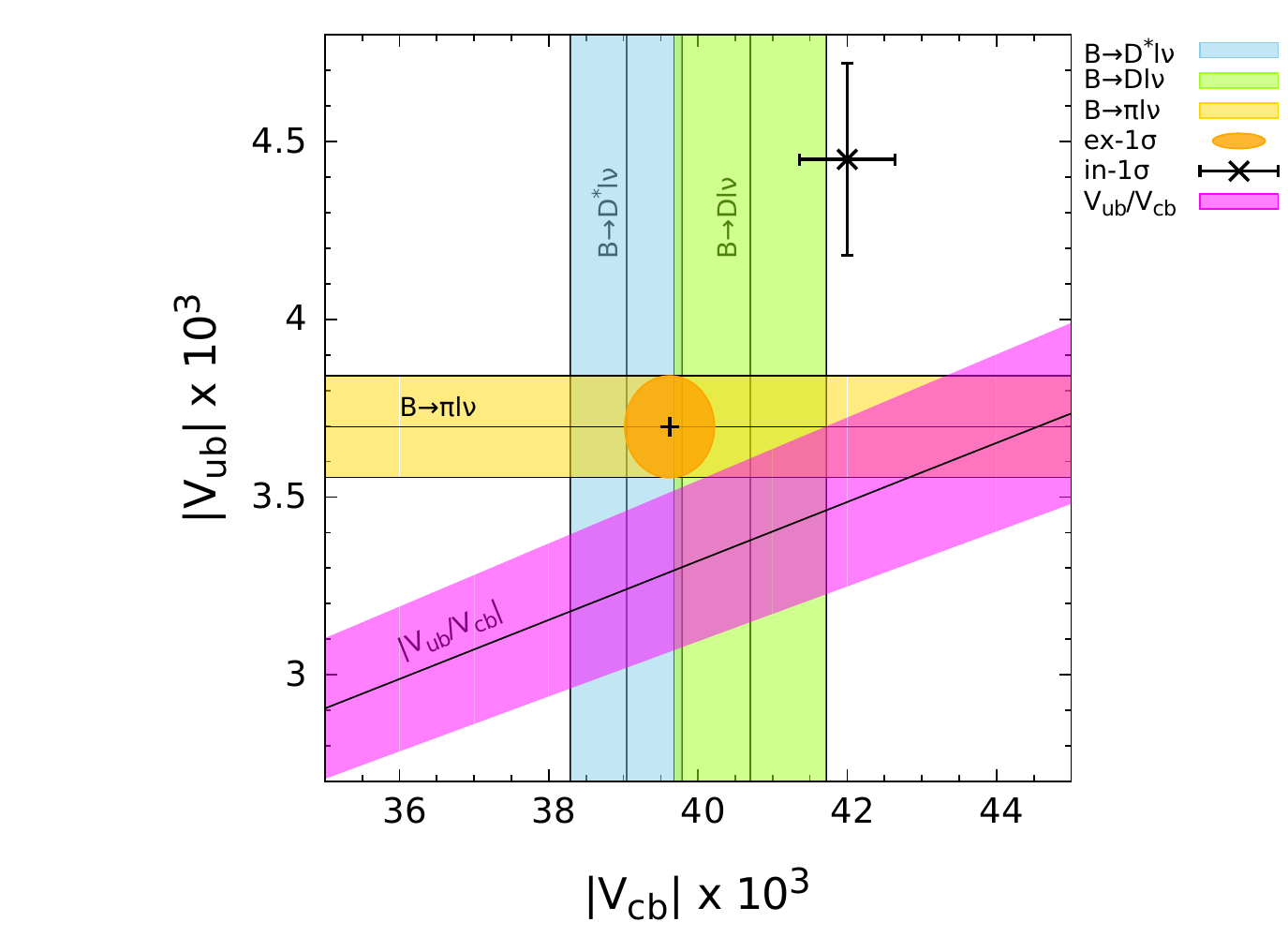}
    \captionof{figure}{$|V_{cb}|$ versus $|V_{ub}|$.}
    \label{fig:Vcb-Vub}
  \end{minipage}
\end{table}
Let us begin with $V_{cb}$.
In Table \ref{tab:Vcb-Vub}, we summarize updated results for $|V_{cb}|$
and $|V_{ub}|$.
In Ref.~\cite{DeTar:2015orc},
DeTar has collected the results for the $\bar{B}\to D\ell\bar{\nu}$
decay mode at non-zero recoil from both lattice QCD \cite{
  Lattice:2015rga, Na:2015kha} and the experiments of Babar \cite{
  Aubert:2008yv} and Belle \cite{ Glattauer:2015yag} to make a
combined fit of all of them.
This result corresponds to the green band in Fig.~\ref{fig:Vcb-Vub}.
We combine the results of Refs.~\cite{DeTar:2015orc} ($\bar{B}\to
D\ell\bar{\nu}$) and \cite{Bailey2014:PhysRevD.89.114504} ($\bar{B}\to
D^*\ell\bar{\nu}$) to obtain the uncorrelated weighted average, which
corresponds to the ``ex-combined'' result in Table \ref{tab:Vcb-Vub}.
This value is shown as an orange circle in Fig.~\ref{fig:Vcb-Vub}.
The black cross represents results of inclusive $|V_{cb}|$ and $|V_{ub}|$.
The inclusive results are about $3\sigma$ away from
those of the exclusive decays as well as 
the LHCb results of $|V_{ub}/V_{cb}|$
(the magenta band in Fig.~\ref{fig:Vcb-Vub}).

\section{Input parameter $\xi_0$}
There are two independent methods to determine $\xi_0$ in lattice QCD:
One is the indirect method, and the other is the direct method.
The parameter $\xi_0$ is connected with $\eps'/\eps$ and $\xi_2$ as follows, 
\begin{align}
&\xi_0  = \frac{\Im A_0}{\Re A_0}, \qquad 
\xi_2 = \frac{\Im A_2}{\Re A_2}, \qquad
\Re \left(\frac{\eps'}{\eps} \right) = 
\frac{\omega}{\sqrt{2} |\eps_K|} (\xi_2 - \xi_0) \,.
\label{eq:e'/e:xi0}
\end{align}
In the indirect method, we determine $\xi_0$ from the experimental
values of $\Re(\eps'/\eps)$, $\eps_K$, $\omega$, and the lattice QCD input
$\xi_2$ using Eq.~\eqref{eq:e'/e:xi0}.
Recently, RBC-UKQCD reported new results for $\xi_2$ in Ref.~\cite{
Blum:2015ywa}.
The results for $\xi_0$ using the indirect method are summarized
in Table \ref{tab:in-LD}.
%
%
% Table of input parameters
%
\begin{table}[t!]
%  \footnotesize
%%%  \renewcommand{\arraystretch}{1.2}
  \renewcommand{\subfigcapskip}{0.55em}
  \subtable[Long Distance Effects]{
    \resizebox{0.47\textwidth}{!}{
      \begin{tabular}{lllc}
        \hline\hline
        Input & Method & Value & Ref. \\ \hline
        $\xi_0$ & indirect & $-1.63(19) \times 10^{-4}$ 
        & \cite{Blum:2015ywa} 
        \\ \hline
        $\xi_0$ & direct  & $-0.57(49) \times 10^{-4}$ 
        & \cite{Bai:2015nea}
        \\ \hline        
        $\xi_\text{LD}$ & --- & $(0 \pm 1.6)\,\%$ 
        & \cite{Christ2012:PhysRevD.88.014508} 
        \\ \hline\hline
      \end{tabular}
    } %%% \resizebox
    \label{tab:in-LD}
  } %%% \subtable
  \hfill
  \subtable[$\BK$]{
    \resizebox{0.47\textwidth}{!}{
      \begin{tabular}{lll}
        \hline\hline
        Collaboration & Value & Ref. 
        \\ \hline
        FLAG-2016   & $0.7625(97)$      & \cite{Aoki:2016frl}
        \\ \hline
        SWME-2014   & $0.7379(47)(365)$ & \cite{Bae2014:prd.89.074504} 
        \\ \hline
        RBC-UK-2016 & $0.7499(24)(150)$ & \cite{Blum:2014tka}
        \\ \hline\hline
      \end{tabular}
    } %%% \resizebox
    \label{tab:in-BK}
  } %%% \subtable
  \caption{ Input parameters: $\xi_0$, $\xi_\text{LD}$ and $\BK$ }
  \label{tab:input-xi}
\end{table}

\begin{figure}[t!]
%  \vspace*{-5mm}
%  \centering
%  \footnotesize
%  \renewcommand{\arraystretch}{1.2}
  \renewcommand{\subfigcapskip}{-0.55em}
  \hspace*{-7mm}
  \subfigure[$\delta_0$]{
    \includegraphics[width=0.55\textwidth]{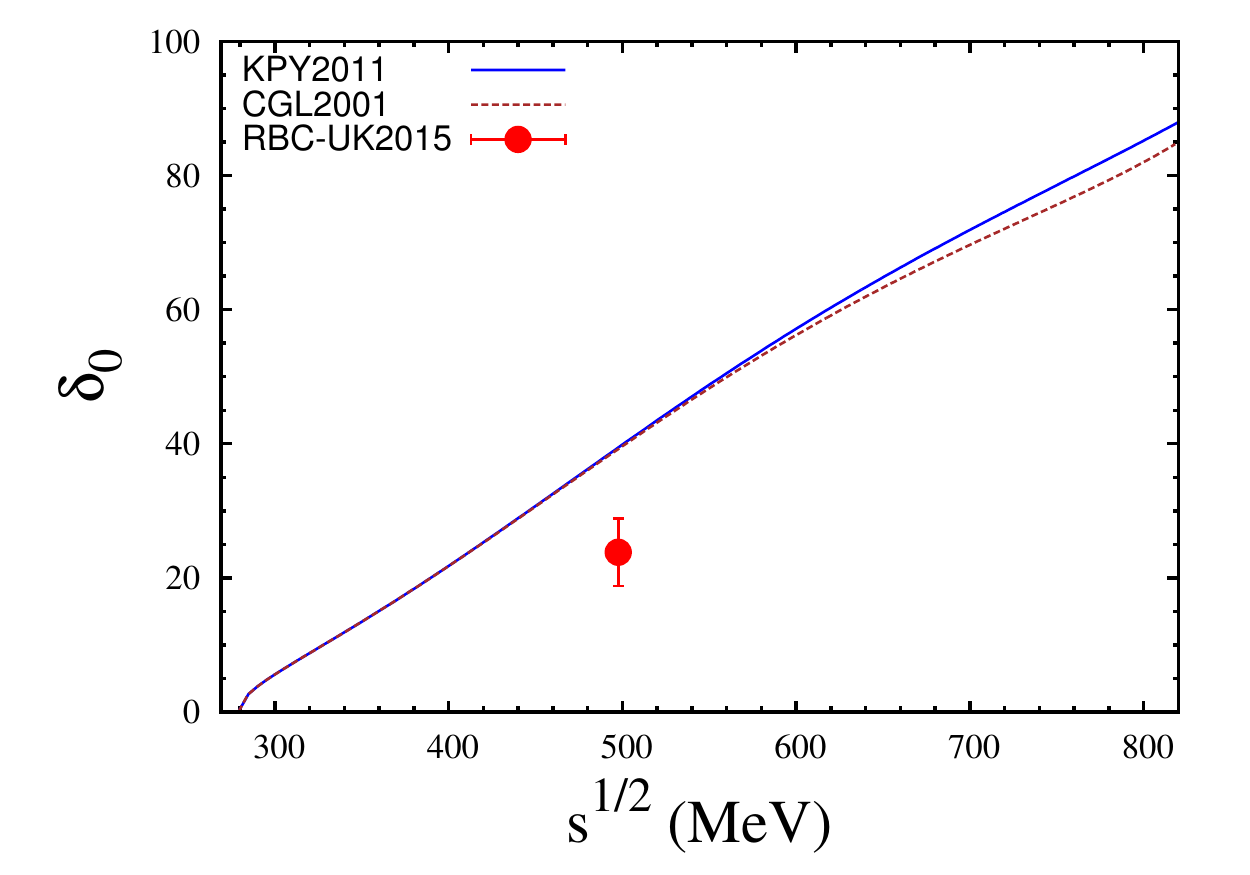}
    \label{fig:d0-rbc}
  }
%%  \hfill
  \hspace*{-7mm}
  \subfigure[{$\delta_2$}]{
    \includegraphics[width=0.55\textwidth]{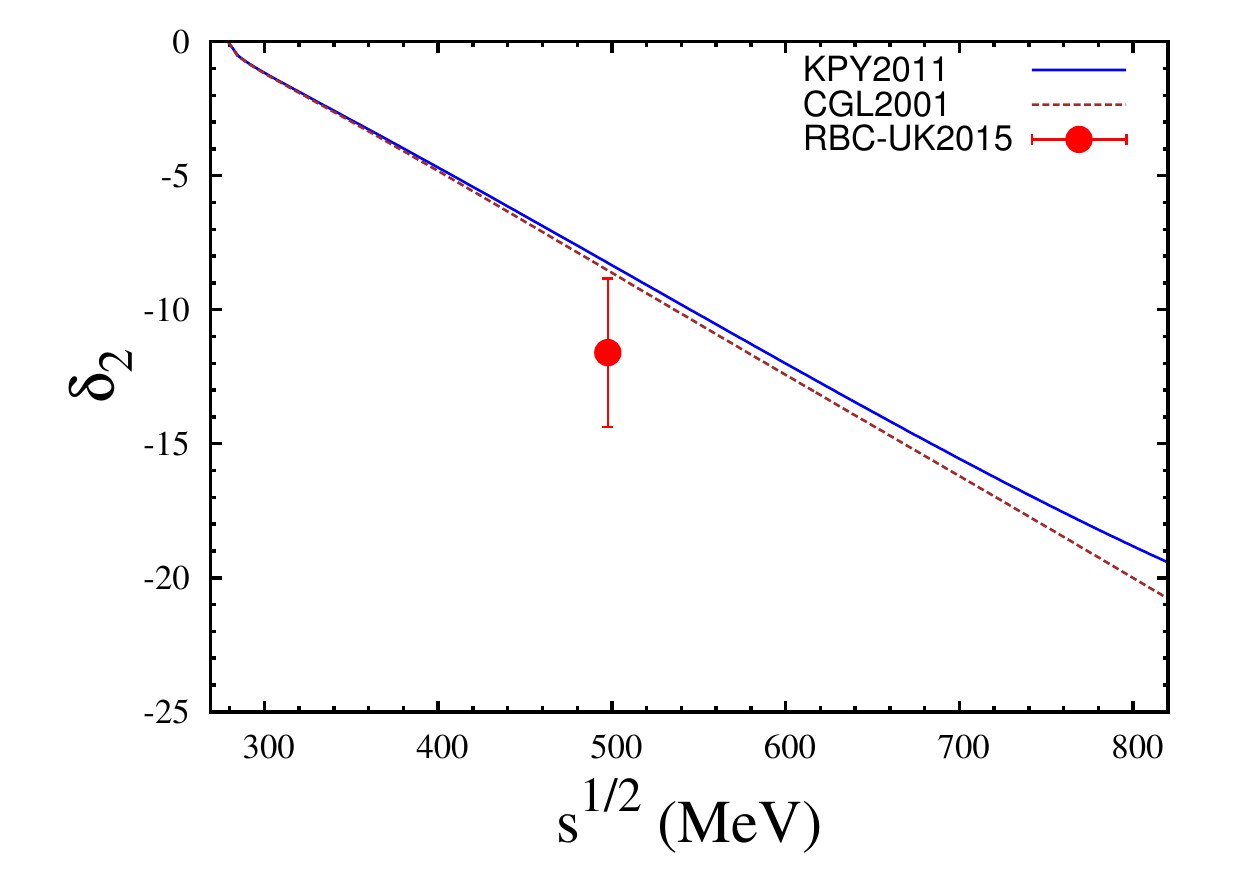}
    \label{fig:d2-rbc}
  }
  \caption{S-wave $\pi-\pi$ scattering phase shifts with $I=0$ and $I=2$.}
  \label{fig:d0-d2}
\end{figure}

Recently, RBC-UKQCD also reported new lattice QCD results for $\Im A_0$
calculated using domain wall fermions \cite{ Bai:2015nea}.
Using the experimental value of $\Re A_0$, we can determine $\xi_0$
directly from $\Im A_0$.
RBC-UKQCD also reported the S-wave $\pi-\pi$ (I=0) scattering
phase shift $\delta_0 = 23.8(49)(12)$ \cite{ Bai:2015nea}.
This value is $3.0\sigma$ lower than the conventional determination
of $\delta_0$ in Refs.~\cite{GarciaMartin:2011cn} (KPY-2011) and
\cite{ Colangelo:2001df, Colangelo2016:MITP} (CGL-2001).
The values for $\delta_0$ are summarized in Table \ref{tab:d0}.
In Fig.~\ref{fig:d0-exp}, we show the results of KPY-2011.
They used a singly subtracted Roy-like equation to do the
interpolation around $\sqrt{s} = m_K$ (kaon mass).
Their fitting to the experimental data works well from the threshold
to $\sqrt{s} = 800\MeV$.
%
%
% Table of input parameters
%
\begin{table}[ht]
  \begin{minipage}[b]{0.45\linewidth}
  \centering
  \renewcommand{\arraystretch}{1.7}
  \resizebox{1.0\textwidth}{!}{
    \begin{tabular}{lll}
      \hline\hline
      Collaboration & $\delta_0$ & Ref. 
      \\ \hline
      RBC-UK-2016  & $23.8(49)(12){}^{\circ}$ & \cite{Bai:2015nea}
      \\ \hline
      KPY-2011 & 39.1(6)${}^{\circ}$ & \cite{GarciaMartin:2011cn}
      \\ \hline
      CGL-2001 & 39.2(15)${}^{\circ}$ & \cite{Colangelo:2001df,Colangelo2016:MITP}
      \\ \hline\hline
    \end{tabular} 
  } %%% \resizebox
  \caption{Results of $\delta_0$}
  \label{tab:d0}
  \end{minipage}
  \hfill
  \begin{minipage}[b]{0.5\linewidth}
    \centering
    \includegraphics[width=\textwidth]{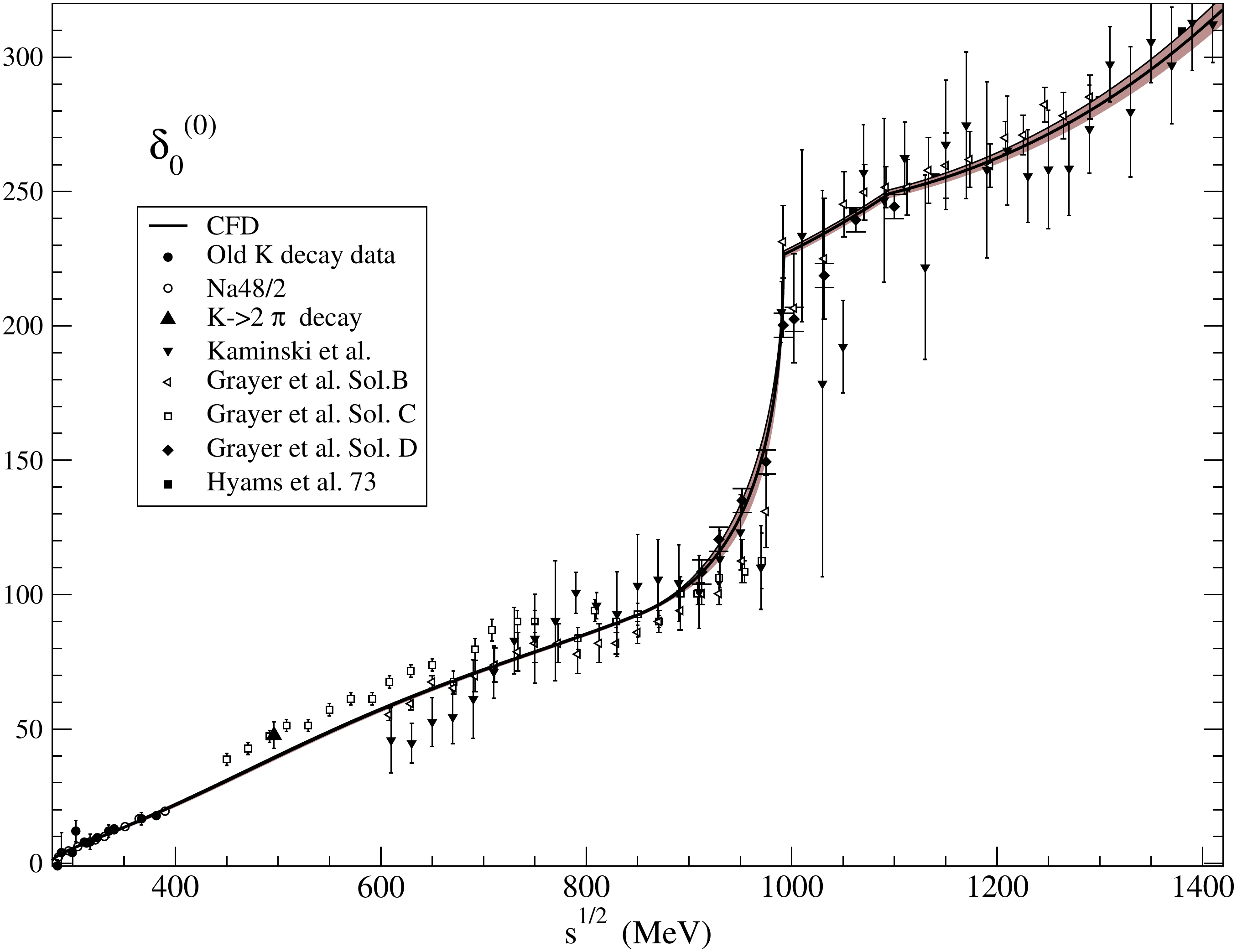}
    \captionof{figure}{Experimental results of $\delta_0$}
    \label{fig:d0-exp}
  \end{minipage}
\end{table}
In Fig.~\ref{fig:d0-rbc}, we show the fitting results of both
KPY-2011 and CGL-2001 as well as the RBC-UKQCD result.
There is essentially no difference between KPY-2011 and CGL-2001 
in the region near $\sqrt{s} = m_K$.
Here, we observe the $3.0\sigma$ gap between RBC-UKQCD and
KPY-2011.
In contrast, in the case of $\delta_2$ (S-wave, I=2), there is no
difference between RBC-UKQCD and KPY-2011 within statistical
uncertainty.
%
%
% Table of input parameters
%

Considering all aspects, we conclude that the direct calculation
of $\Im A_0$ and $\xi_0$ by RBC-UKQCD in Ref.~\cite{ Bai:2015nea}
may have unresolved issues.
Hence, we use the indirect method to determine $\xi_0$ in this paper.

Regarding $\xi_\text{LD}$, the long distance effect in
the dispersive part, there has been an on-going attempt to calculate
it on the lattice \cite{ Christ:2014qwa}.
However, this attempt \cite{Bai2016:Latt}, at present,
belongs to the category of
exploratory study rather than to that of precision measurement.
Hence, we use the rough estimate of $\xi_\text{LD}$ in Ref.~\cite{
Christ:2014qwa} in this paper, which is given in Table
  \ref{tab:in-LD}.

\section{Input parameter $\BK$}
In Table \ref{tab:in-BK}, we present results for $\BK$ calculated in
lattice QCD with $N_f=2+1$ flavors.
Here, FLAG-2016 represents the global average over the results of
BMW-2011 \cite{ Durr2011:PhysLettB.705.477}, Laiho-2011 \cite{
  Laiho:2011np}, RBC-UK-2016 \cite{ Blum:2014tka}, and SWME-2016
\cite{ Jang:2015sla}, which is reported in Ref.~\cite{ Aoki:2016frl}.
SWME-2014 represents the $\BK$ result reported in Ref.~\cite{
Bae2014:prd.89.074504}.
RBC-UK-2016 represents that reported in Ref.~\cite{ Blum:2014tka}.

The results of SWME-2016 are obtained using fitting based on staggered
chiral perturbation theory (SChPT) in the infinite volume limit, while
those of SWME-2014 are obtained using fitting based on SChPT with
finite volume corrections included at the NLO level.
In this paper, we use the FLAG-2016 result of $\BK$.

\section{Other input parameters}
For the Wolfenstein parameters $\lambda$, $\bar{\rho}$, and
$\bar{\eta}$, both CKMfitter and UTfit updated their results in
Refs.~\cite{ Charles:2015gya, UTfit2016:web}, while the angle-only-fit
has not been updated since 2015.
The results are summarized in Table \ref{tab:in-wolf}.

For the QCD corrections $\eta_{cc}$, $\eta_{ct}$, and
$\eta_{tt}$, we use the same values as in Ref.~\cite{ Bailey:2015tba},
which are given in Table \ref{tab:in-eta}.
Other input parameters are the same as in Ref.~\cite{
Bailey:2015tba} except for the charm quark mass $m_c(m_c)$, which are
  summarized in Table \ref{tab:in-extra}.
For the charm quark mass, we use the HPQCD results of $m_c(m_c)$
  reported in Ref.~\cite{ Chakraborty:2014aca}.
%
%
% Table of input parameters
%
\begin{table}[t!]
  \centering
  \renewcommand{\subfigcapskip}{0.55em}
  \subtable[Wolfenstein parameters]{
    \resizebox{0.63\textwidth}{!}{
      \begin{tabular}{cccc}
        \hline\hline
        & CKMfitter & UTfit & AOF \cite{Bevan2013:npps241.89} \\ \hline
        $\lambda$
        & $0.22548(68)$
        /\cite{Charles:2015gya}
        & $0.22497(69)$
        /\cite{UTfit2016:web}
        & $0.2253(8)$
        /\cite{Agashe2014:ChinPhysC.38.090001}
        \\ \hline
        $\bar{\rho}$
        & $0.145(13)$
        /\cite{Charles:2015gya}
        & $0.153(13)$
        /\cite{UTfit2016:web}
        & $0.139(29)$
        /\cite{UTfit2014PostMoriondSM:web}
        \\ \hline
        $\bar{\eta}$
        & $0.343(12)$
        /\cite{Charles:2015gya}
        & $0.343(11)$
        /\cite{UTfit2016:web}
        & $0.337(16)$
        /\cite{UTfit2014PostMoriondSM:web}
        \\ \hline\hline
      \end{tabular} 
    } %%% \resizebox
    \label{tab:in-wolf}
  } %%% \subtable
  \\
  \subtable[QCD corrections]{
    \resizebox{0.32\textwidth}{!}{
      \begin{tabular}{clc}
        \hline\hline
        Input & Value & Ref. \\ \hline
        $\eta_{cc}$ & $1.72(27)$
        & { \protect\cite{Bailey:2015tba} } \\ \hline
        $\eta_{tt}$ & $0.5765(65)$
        & { \protect\cite{Buras2008:PhysRevD.78.033005} } \\ \hline
        $\eta_{ct}$ & $0.496(47)$
        & { \protect\cite{Brod2010:prd.82.094026} }
        \\ \hline\hline
      \end{tabular}
      } %%% \resizebox
    \label{tab:in-eta}
  } %%% \subtable
  \hfill
  \subtable[Other input parameters]{
    \resizebox{0.5\textwidth}{!}{
      \begin{tabular}{clc}
        \hline\hline
        Input & Value & Ref. \\ \hline
        $G_{F}$
        & $1.1663787(6) \times 10^{-5}$ GeV$^{-2}$
        &\cite{Agashe2014:ChinPhysC.38.090001} \\ \hline
        $M_{W}$
        & $80.385(15)$ GeV
        &\cite{Agashe2014:ChinPhysC.38.090001} \\ \hline
        $m_{c}(m_{c})$
        & $1.2733(76)$ GeV
        &\cite{Chakraborty:2014aca} \\ \hline
        $m_{t}(m_{t})$
        & $163.3(2.7)$ GeV
        &\cite{Alekhin2012:plb.716.214} \\ \hline
        $\theta$
        & $43.52(5)^{\circ}$
        &\cite{Agashe2014:ChinPhysC.38.090001} \\ \hline
        $m_{K^{0}}$
        & $497.614(24)$ MeV
        &\cite{Agashe2014:ChinPhysC.38.090001} \\ \hline
        $\Delta M_{K}$
        & $3.484(6) \times 10^{-12}$ MeV
        &\cite{Agashe2014:ChinPhysC.38.090001} \\ \hline
        $F_K$
        & $156.2(7)$ MeV
        &\cite{Agashe2014:ChinPhysC.38.090001}
        \\ \hline\hline
      \end{tabular}
    } %%% \resizebox
    \label{tab:in-extra}
  } %%% \subtable
  \caption{ Input parameters }
  \label{tab:input}
\end{table}
%

%%% EDIT-jon-wlee

\section{Results for $\epsK$ with lattice QCD inputs}
\begin{figure}[t!]
%  \vspace*{-5mm}
%  \centering
%  \footnotesize
%  \renewcommand{\arraystretch}{1.2}
%  \renewcommand{\subfigcapskip}{-0.55em}
%
  \hspace*{-7mm}
  \subfigure[$\epsK$ with exclusive $V_{cb}$]{
    \includegraphics[width=0.50\textwidth]{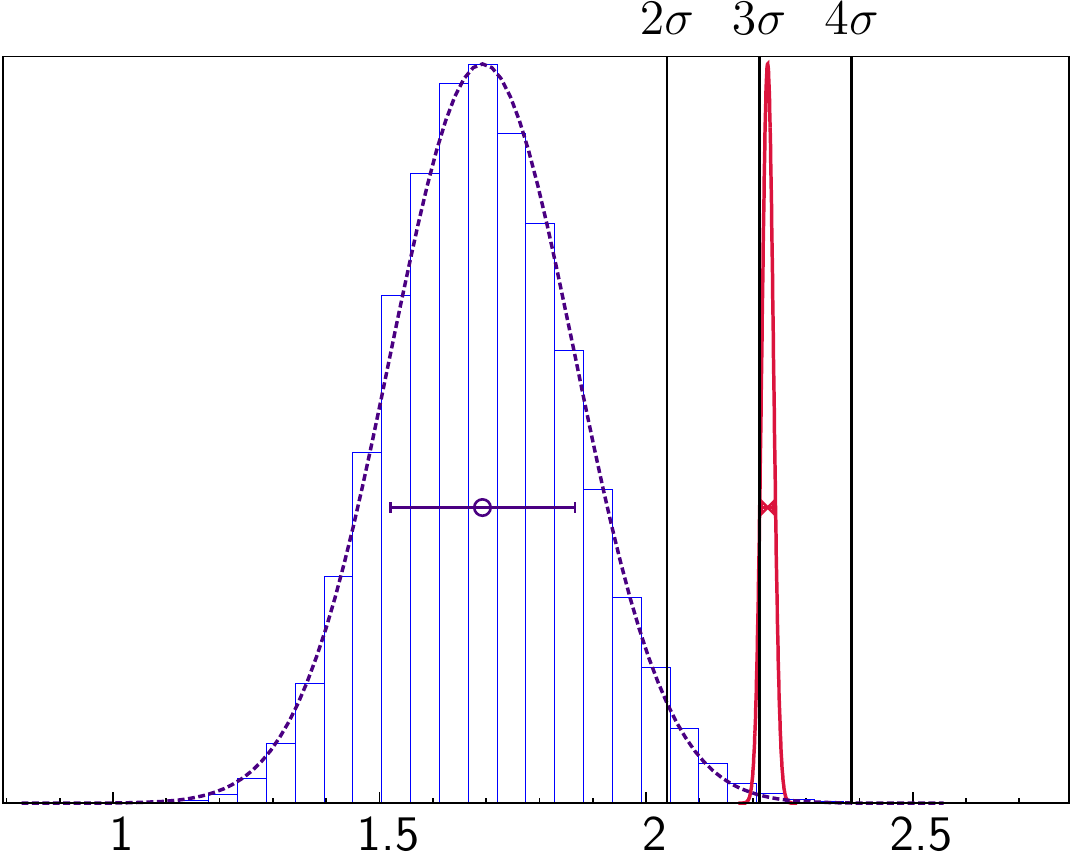}
    \label{fig:eps-flag-ex}
  }
  \hfill
  \hspace*{-7mm}
  \subfigure[{$\epsK$} with inclusive $V_{cb}$]{
    \includegraphics[width=0.50\textwidth]{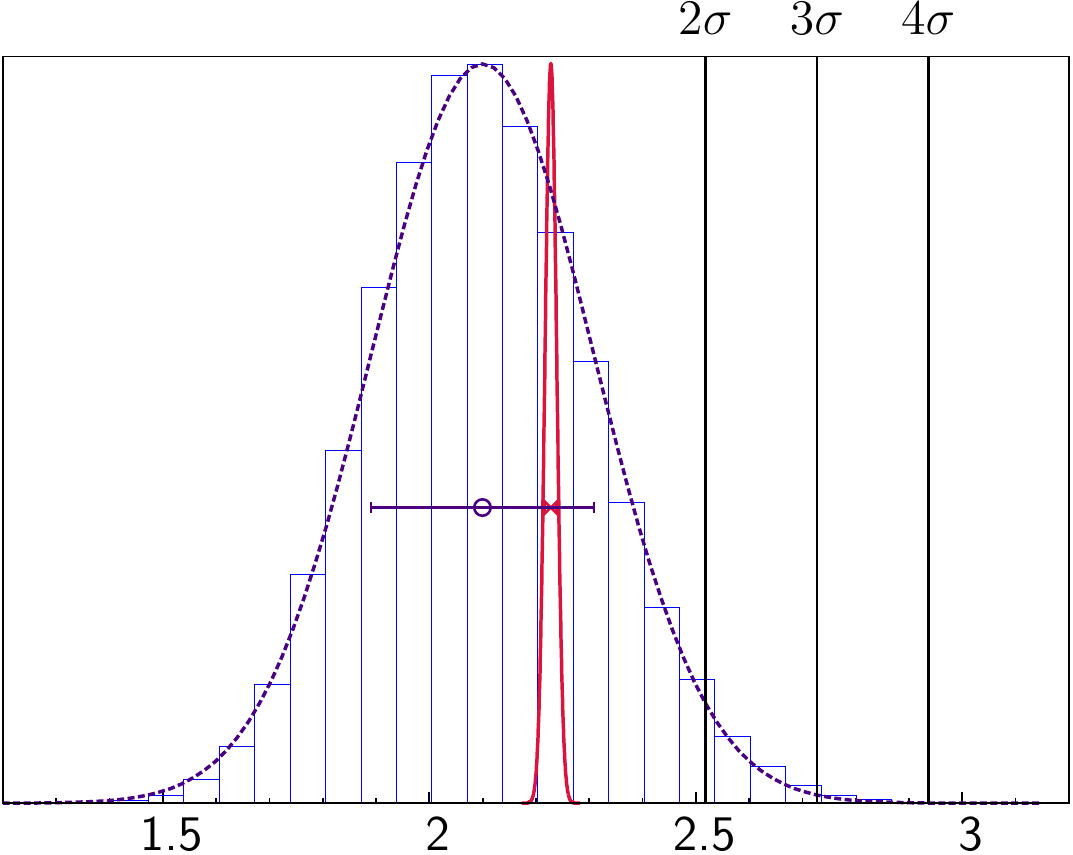}
    \label{fig:eps-flag-in}
  }
  \caption{$\epsK$ with exclusive $V_{cb}$ (left) and inclusive
    $V_{cb}$ (right). Here, we use the FLAG-2016 $\BK$ and AOF for the
    Wolfenstein parameters. The red curve represents the experimental
    value of $\epsK$ and the blue curve the theoretical value evaluated
    directly from the standard model.}
  \label{fig:eps-flag}
\end{figure}

In Fig.~\ref{fig:eps-flag}, we show the results for $\epsK$ evaluated
directly from the standard model with the lattice QCD inputs described
in the previous sections.
In Fig.~\ref{fig:eps-flag-ex}, the blue curve represents the
theoretical evaluation of $\epsK$ with the FLAG $\BK$, AOF for the
Wolfenstein parameters, and exclusive $V_{cb}$ that corresponds to 
ex-combined in Table \ref{tab:Vcb-Vub}.
Here the red curve represents the experimental value of $\epsK$.
In Fig.~\ref{fig:eps-flag-in}, the blue curve represents the same as
in \ref{fig:eps-flag-ex} except for using the inclusive $V_{cb}$
in Table \ref{tab:Vcb-Vub}.
Our preliminary results are, in units of $1.0\times 10^{-3}$,
\begin{align}
  |\epsK| &= 1.69 \pm 0.17 && \text{for exclusive $V_{cb}$ (lattice QCD)}
  \\
  |\epsK| &= 2.10 \pm 0.21 && \text{for inclusive $V_{cb}$ (QCD sum rules)}
  \\
  |\epsK| &= 2.228 \pm 0.011 && \text{(experimental value)}
\end{align}
This indicates that there is $3.2\sigma$ tension in the exclusive
$V_{cb}$ channel (lattice QCD) and no tension in the inclusive
$V_{cb}$ channel (heavy quark expansion; QCD sum rules).
%
%
%
% Table of input parameters
%

\acknowledgments
We thank R.~Van de Water for helpful discussion on $V_{cb}$.
The research of W.~Lee is supported by the Creative Research
Initiatives Program (No.~20160004939) of the NRF grant funded by the
Korean government (MEST).
~J.A.B. is supported by the Basic Science Research Program of the
National Research Foundation of Korea (NRF) funded by the Ministry of
Education (No.~2015024974).
W.~Lee would like to acknowledge the support from the KISTI
supercomputing center through the strategic support program for the
supercomputing application research (No.~KSC-2014-G3-003).
Computations were carried out on the DAVID GPU clusters at Seoul
National University.
%

% EDIT

%------------
% references
%------------
\bibliography{refs}

\end{document}